\documentclass{article}

\usepackage[accepted]{icml2025}
\setcitestyle{numbers,sort&compress}

\usepackage{microtype}
\usepackage{graphicx}
\usepackage{enumitem}
\graphicspath{{Figures/}}
\usepackage{subfigure}
\usepackage{booktabs}
\usepackage{amsmath, amssymb, mathtools}
\usepackage{amsthm}
\usepackage[colorlinks=true,linkcolor=black,citecolor=blue,urlcolor=blue]{hyperref}

\usepackage[capitalize,noabbrev]{cleveref}

\usepackage{float}
\usepackage{placeins}

\usepackage{algorithm}

\theoremstyle{plain}

\theoremstyle{definition}

\theoremstyle{remark}

\icmltitlerunning{Speculative Decoding in Decentralized LLM Inference}

\makeatletter
\def\ICML@appearing{}  
\makeatother

\begin{document}

\twocolumn[

{\raggedright
  \vspace{-0.2cm} 
  \includegraphics[height=1.2cm]{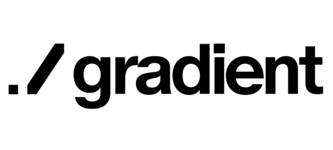}\par\vspace{-0.3cm} 
}

\icmltitle{Speculative Decoding in Decentralized LLM Inference:\\
Turning Communication Latency into Computation Throughput}

\begin{center}
\textbf{Jingwei Song$^{1,2}$, Wanyi Chen$^{1,3}$, Xinyuan Song$^{1,4}$,Max$^{1}$,Chris Tong$^{1}$,Gufeng Chen$^{1}$ ,Tianyi Zhao$^{1}$ ,Eric Yang$^{1}$, Bill Shi$^{1\dagger}$, Lynn Ai$^{1}$ }
\par\medskip
$^{1}$Gradient Network
\par
$^{2}$The University of Hong Kong
\par
$^{3}$Soochow University
\par
$^{4}$Emory University

\end{center}
% -----------------------------------------------------------------
\icmlcorrespondingauthor{Bill Shi}{tianyu@gradient.network}
\vskip 0.3in
]

\printAffiliationsAndNotice{}        

\begin{abstract}
Speculative decoding accelerates large language model (LLM) inference by using a lightweight draft model to propose tokens that are later verified by a stronger target model. While effective in centralized systems, its behavior in decentralized settings, where network latency often dominates compute, remains under-characterized. We present \textbf{Decentralized Speculative Decoding (DSD)}, a plug and play framework for decentralized inference that turns communication delay into useful computation by verifying multiple candidate tokens in parallel across distributed nodes. We further introduce an adaptive speculative verification strategy that adjusts acceptance thresholds based on token level semantic importance and, in our experiments, provides up to an additional 15\% to 20\% end to end speedup over a nonadaptive speculative decoding baseline, without retraining. In theory, DSD reduces cross node communication cost by approximately $(N-1)t_{1}(k-1)/k$, where $t_{1}$ is per link latency and $k$ is the average number of tokens accepted per round. In practice, DSD achieves up to $2.56\times$ speedup on HumanEval and $2.59\times$ on GSM8K, surpassing the Eagle3 baseline while preserving accuracy. These results show that adapting speculative decoding for decentralized execution provides a system level optimization that converts network stalls into throughput, enabling faster distributed LLM inference with no model retraining or architectural changes.
\end{abstract}

\section{Introduction}

Large language models (LLMs) continue to grow in size, and inference efficiency has become a critical concern for both research and production systems~\cite{brown2020gpt3,kaplan2020scaling,hoffmann2022chinchilla,narayanan2021efficient}. While modern accelerators provide increasing computational power, performance bottlenecks have shifted toward memory bandwidth and, in distributed settings, inter node communication latency~\cite{narayanan2021efficient,rajbhandari2021deepspeed,shoeybi2019megatron,huang2019gpipe}. Many existing efficiency techniques such as quantization~\cite{dettmers2022llmint8}, tensor parallelism~\cite{shazeer2019mesh,rajbhandari2021deepspeed}, and speculative decoding~\cite{leviathan2023fastinferencetransformersspeculative} were developed for centralized or single server settings. As decentralized inference frameworks such as \textbf{Parallax}~\cite{tong2025parallaxefficientllminference} become more common and execution spans geographically distributed nodes, communication latency can exceed per step computation time, which severely reduces the benefits of classical methods designed for centralized computation~\cite{narayanan2021efficient,rajbhandari2021deepspeed,shoeybi2019megatron}. In this work we follow Parallax and use the term \emph{decentralized inference} to refer to LLM serving over multiple independently managed nodes that may reside in different regions.

Speculative decoding~\cite{leviathan2023fastinferencetransformersspeculative,Miao_2024} accelerates autoregressive generation by pairing a lightweight draft model with a stronger target model. The draft model predicts several candidate tokens, and the target model verifies them in a single forward pass, reducing the number of target model evaluations. Prior studies typically assume that computation dominates the total cost~\cite{chen2023specinfer,li2023eagle}. In decentralized settings, however, this assumption no longer holds, because cross node communication becomes the primary overhead~\cite{narayanan2021efficient}.

We revisit speculative decoding from a distributed systems perspective and propose \textbf{Decentralized Speculative Decoding (DSD)}, a communication aware formulation that converts network waiting time into useful computation. Instead of leaving nodes idle during communication, DSD verifies $k$ tokens jointly across distributed nodes, reducing synchronization from $k$ rounds to a single round. For a system with $N$ nodes, per link latency $t_1$, and local compute time $t_0$, this reduces communication latency by approximately $(N-1)t_1(k-1)/k$. The advantage is most pronounced when $3 \le N \le 8$ and $3t_0 < t_1 < 10t_0$, a regime commonly observed in wide area or mixed hardware deployments~\cite{rajbhandari2021deepspeed}.

In addition, we introduce an \textbf{adaptive speculative verification} strategy that does not require retraining. Instead of applying the same acceptance rule to all tokens, DSD identifies \emph{high impact tokens} using cross entropy contrast, token match statistics, and a distributional agreement score, and applies a relaxation factor $\tau$ to adjust acceptance strength. Tokens with strong semantic influence are verified strictly, while low impact tokens are accepted more flexibly. In our experiments, this inference only mechanism provides up to an additional 15\% to 20\% end to end acceleration over a nonadaptive speculative decoding baseline while preserving output quality.

We implement DSD in the \emph{Parallax} decentralized inference engine~\cite{tong2025parallaxefficientllminference} and evaluate it on HumanEval, GSM8K, Alpaca, MT Bench, and CNN/DailyMail using \texttt{Llama3.1-8B} and \texttt{Qwen3-8B}. DSD achieves up to \textbf{2.56$\times$} speedup on HumanEval and \textbf{2.59$\times$} on GSM8K, matching or exceeding the Eagle3 baseline~\cite{li2023eagle} without accuracy degradation.

Our contributions are summarized as follows:
\begin{itemize}[left=0em]
\item We formulate speculative decoding for decentralized inference, reducing communication frequency by verifying multiple tokens in each synchronization round without modifying model weights.
\item We propose a training free adaptive verification method that adjusts token level acceptance according to semantic relevance, increasing accepted token spans while preserving accuracy.
\item We provide a system level analysis that connects communication latency and verification efficiency, supported by experiments on reasoning and code generation benchmarks that show consistent acceleration at accuracy parity.
\end{itemize}

\section{Methodology}
\label{sec:method}

We formalize DSD and describe two components: (i) speculative parallelism adapted to decentralized inference, and (ii) adaptive speculative verification that is semantically aware and training free. % semantic aware -> semantically aware
The objective is to increase Model FLOPs Utilization (MFU) and reduce inter node latency without retraining or modifying model weights, consistent with recent system level optimization efforts~\cite{narayanan2021efficient,rajbhandari2021deepspeed}.

\subsection{Preliminaries: Standard Speculative Decoding}
\label{sec:prelim}

\begin{figure}[!ht]
\centering
\includegraphics[width=\linewidth]{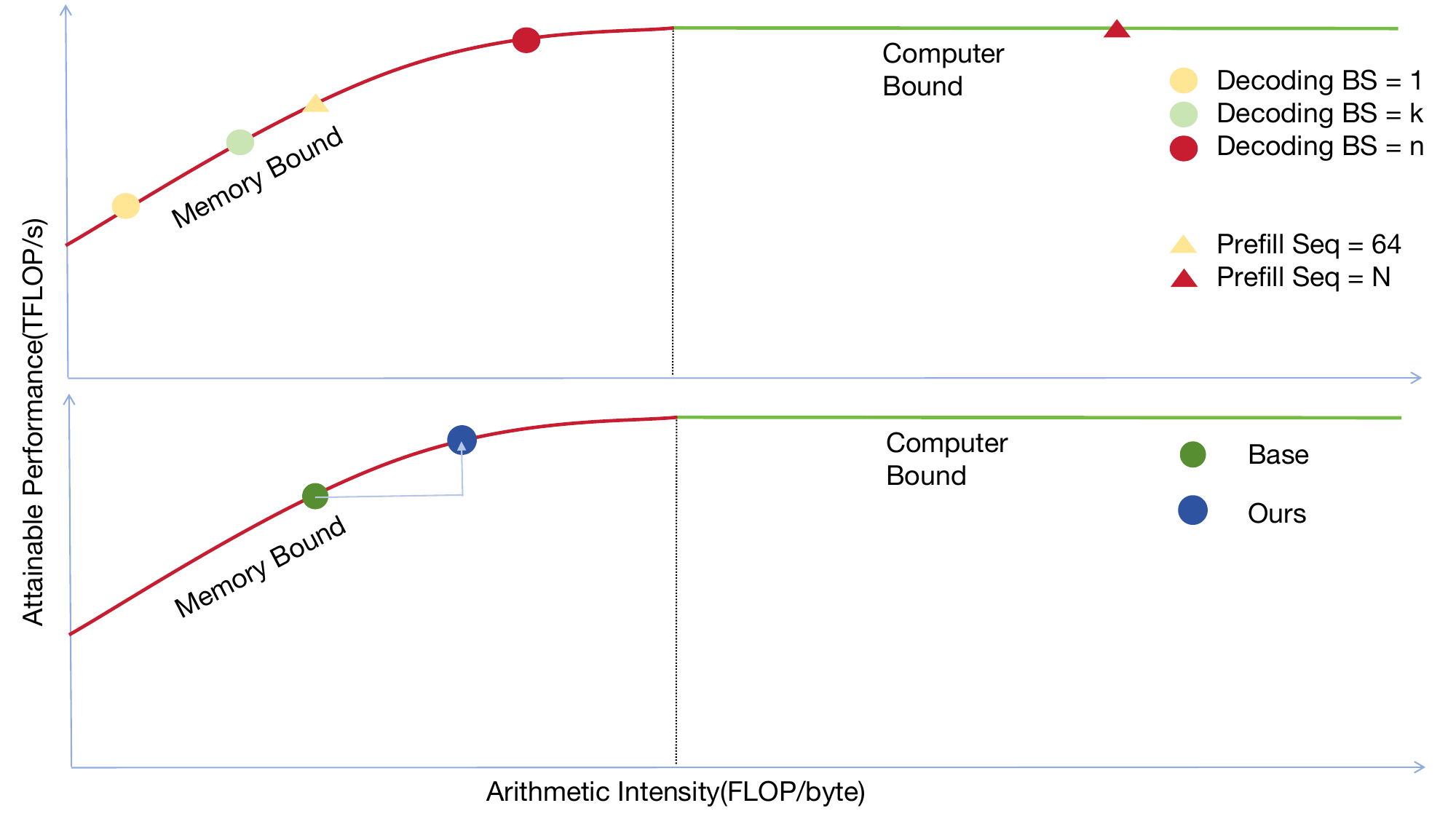}
\caption{Roofline view of attainable performance versus arithmetic intensity. Prefill is compute oriented at moderate batch size, while token by token decoding is often memory bound~\cite{dao2022flashattention}. Verifying a compact draft window increases effective intensity.}
\label{fig:roofline}
\end{figure}

Speculative decoding~\cite{leviathan2023fastinferencetransformersspeculative,Miao_2024} accelerates autoregressive generation by pairing a lightweight \emph{draft model} $M_d$ with a higher accuracy \emph{target model} $M_t$. Given context $x_{1:i}$, the draft produces a window of $\gamma$ candidate tokens
\begin{equation}
\hat{y}_{i+1:i+\gamma} = M_d(x_{1:i}),
\end{equation}
which the target verifies in a single forward pass. Let $k \le \gamma$ be the number of accepted tokens. The sequence advances by appending the first $k$ accepted tokens and then sampling one additional token from $M_t$:
\begin{equation}
x_{1:i+k} = x_{1:i} \oplus \hat{y}_{i+1:i+k}.
\end{equation}
Processing a window of $\gamma$ tokens at once raises effective throughput by roughly $(\gamma+1)$ per target evaluation~\cite{li2023eagle,chen2023specinfer}. Figure~\ref{fig:roofline} illustrates how verifying a compact window shifts decoding toward a region with higher arithmetic intensity.

\subsection{Decentralized Inference Model}
\label{sec:decentralized-model}

Consider $N$ participating nodes, each hosting a model shard as in pipeline or tensor parallel inference systems~\cite{shoeybi2019megatron,huang2019gpipe}. Let $t_0$ denote the local compute time per decoding step and $t_1$ the point to point communication latency. With standard autoregressive decoding, producing each token requires synchronization across nodes, yielding communication overhead $(N-1)t_1$ per token. The time to generate $k$ tokens is
\begin{equation}
T_{\text{std}} = k\bigl(t_0 + (N-1)t_1\bigr).
\end{equation}

\begin{figure}[!ht]
\centering
\includegraphics[width=\linewidth]{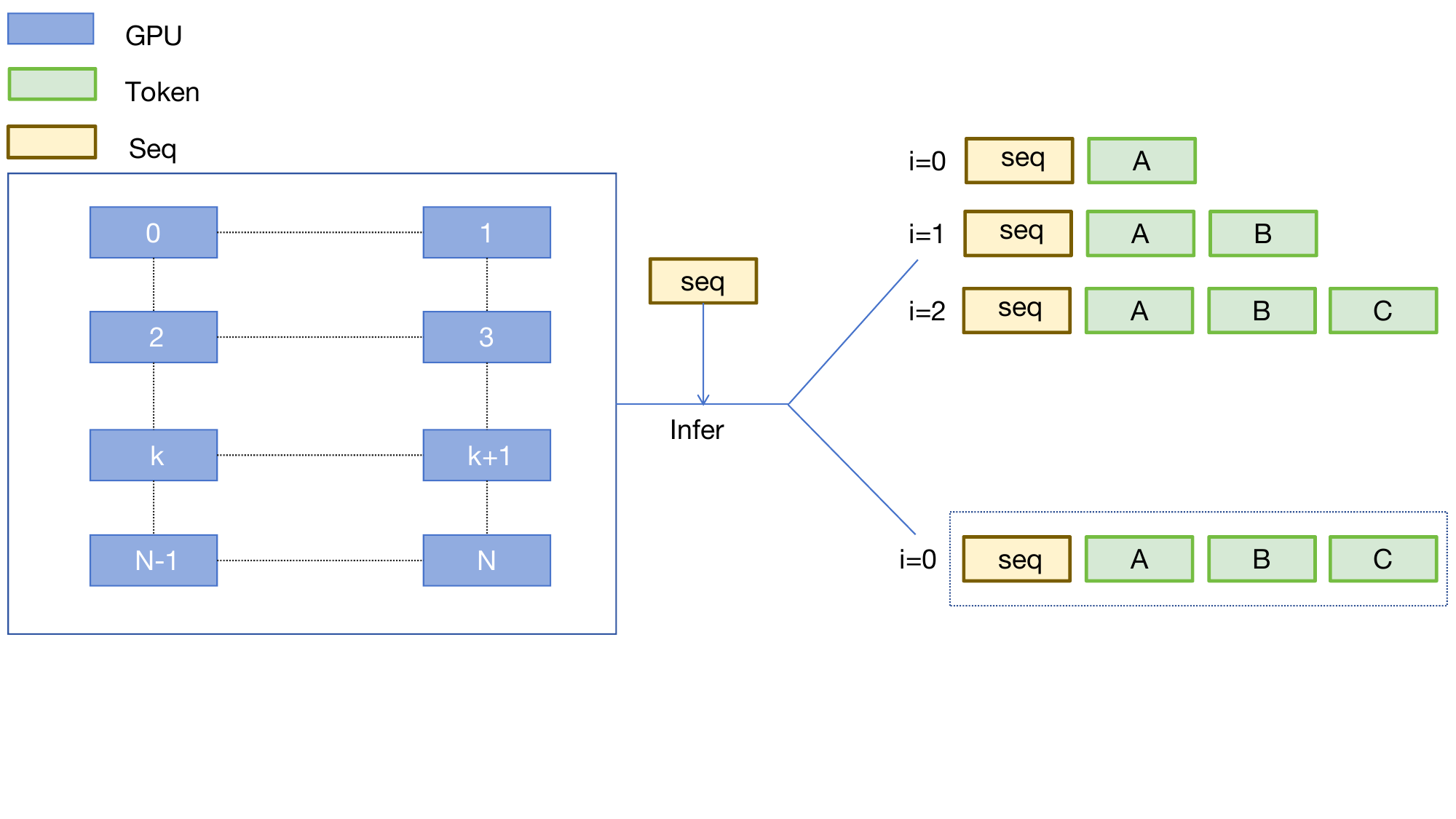}
\caption{Decentralized speculative decoding. Left: model shards on distributed nodes. Right: a draft window is verified; if the first $k$ tokens pass, they are committed together in one round, improving pipeline utilization.}
\label{fig:system}
\end{figure}

With decentralized speculative decoding (DSD), $k$ speculative tokens are verified in one synchronization round, which amortizes communication across the window:
\begin{equation}
T_{\text{DSD}} = k\,t_0 + (N-1)t_1.
\end{equation}
The communication reduction ratio is therefore
\begin{equation}
R_{\text{comm}}
= 1 - \frac{T_{\text{DSD}}}{T_{\text{std}}}
= \frac{(N-1)t_1\, (k-1)}{k\bigl(t_0 + (N-1)t_1\bigr)}.
\end{equation}
The benefit is most pronounced when $3 \le N \le 8$ and $3t_0 < t_1 < 10t_0$, a range commonly observed in wide area or mixed hardware deployments~\cite{tong2025parallaxefficientllminference}. Figure~\ref{fig:system} depicts the runtime: a draft window is verified and, if the first $k$ tokens pass, they are committed in a single synchronization, converting waiting into useful computation.

\subsection{Adaptive Speculative Verification}
\label{sec:adaptive-verification}

Classical speculative decoding enforces strict token level distribution consistency,
\begin{equation}
P_t(y_i \mid x_{1:i-1}) = P_d(y_i \mid x_{1:i-1}),
\end{equation}
which ensures accuracy but can limit acceptance due to tight matching~\cite{leviathan2023fastinferencetransformersspeculative}. DSD adopts adaptive verification: tokens with low semantic effect receive relaxed checks, while tokens that are expected to influence future semantics are verified rigorously.

\paragraph{Key token identification.}
A token $y_i$ is labeled as a key token if any of the following holds:
\begin{equation}
\begin{aligned}
\mathrm{Key}(y_i)\iff\;&
\Bigl[\tfrac{H_d(y_i)}{H_t(y_i)}>\lambda_1\Bigr]
\;\lor\;
\Bigl[\lvert P_t(y_i)-P_d(y_i)\rvert>\lambda_2\Bigr]\\
&\lor\;
\Bigl[\mathrm{NormMatch}(y_i)<\lambda_3\Bigr],
\end{aligned}
\label{eq:key}
\end{equation}
where $H_d$ and $H_t$ denote draft and target token cross entropies, and $\mathrm{NormMatch}(y_i)\in[0,1]$ measures normalized distribution similarity between the two predictive distributions (for example based on the overlap of their top-$k$ support). % 多解释半句 NormMatch

Thresholds $\lambda_1,\lambda_2,\lambda_3$ are calibrated on a small validation set.

\paragraph{Verification scaling.}
For non key tokens, a softened target distribution is formed as
\begin{equation}
\tilde{P}_t(y_i) \coloneqq P_t(y_i)^{1-\tau}\, P_d(y_i)^{\tau}, \qquad \tau \in [0,1],
\end{equation}
and then renormalized to obtain a probability distribution over the vocabulary. % 补一句归一化
Acceptance is applied against $\tilde{P}_t$. Key tokens always use $\tau=0$ for strict verification. Modest values of $\tau$ extend accepted spans while preserving alignment with the target model.

\begin{algorithm}[!ht]
\caption{Decentralized Speculative Decoding (DSD)}
\label{alg:dsd}
\begin{algorithmic}[1]
\REQUIRE Draft model $M_d$, target model $M_t$, context $x_{1:i}$, window $\gamma$, relaxation $\tau$
\STATE $\hat{y}_{i+1:i+\gamma} \gets M_d(x_{1:i})$
\STATE Compute criteria and obtain key token set $\mathcal{K}$
\STATE $k \gets 0$
\FOR{each $y_j \in \hat{y}_{i+1:i+\gamma}$ in order}
  \IF{$y_j \in \mathcal{K}$}
    \STATE verify using $P_t$
  \ELSE
    \STATE form $\tilde{P}_t$ and verify
  \ENDIF
  \IF{accepted}
    \STATE append $y_j$; $k \gets k+1$
  \ELSE
    \STATE \textbf{break}
  \ENDIF
\ENDFOR
\STATE sample one additional token from $M_t$
\STATE broadcast $k$ accepted tokens in a single synchronization round
\end{algorithmic}
\end{algorithm}

\subsection{Complexity and Scalability}
\label{sec:complexity}

Let $\rho = k/(\gamma+1)$ denote the mean acceptance ratio. The expected speedup over standard decoding is
\begin{equation}
S = \frac{t_0 + (N-1)t_1}{t_0/\rho + (N-1)t_1/k}.
\end{equation}
DSD improves end to end throughput through two complementary effects: longer accepted spans increase compute utilization, and amortized synchronization reduces communication cost. In practice, $\gamma$ is set to balance arithmetic intensity and acceptance stability, and $\tau$ is chosen in a modest range (for example $0.1$ to $0.3$) to extend acceptance without harming accuracy~\cite{li2023eagle}. The additional statistics for adaptive verification (cross entropies and NormMatch) are computed from logits and add only a small constant factor overhead on top of standard decoding. These settings allow DSD to maintain consistent quality while scaling to distributed and cross region clusters.

\section{Experiments}
\label{sec:experiments}

We evaluate DSD across multiple model sizes, datasets, and deployment settings. The analysis focuses on three questions: (i) how much acceleration DSD achieves relative to centralized speculative decoding and strong systems such as Eagle3~\cite{li2023eagle}, (ii) whether adaptive verification maintains accuracy across varied benchmarks, and (iii) how performance changes as the accepted span $k$ and the relaxation coefficient $\tau$ vary.

\begin{table*}[!ht]
\centering
\caption{Results and ablations across datasets and parameter settings.}
\label{tab:results_main}

\begingroup
\footnotesize
\renewcommand{\arraystretch}{0.85}
\setlength{\tabcolsep}{3pt}

\begin{tabular}{llcccc}
\toprule
\textbf{Target Model} & \textbf{Variable} & \textbf{Base Acc} & \textbf{Eagle3 Acc} & \textbf{Speedup ($\times$)} & \textbf{Avg len} \\
\midrule
\multicolumn{6}{c}{\textit{HumanEval (Llama3.1-8B)}} \\
\midrule
$t{=}0.0,\; qx{=}1$ & -- & 0.6707 & 0.6707 & 4.80 & 5.93 \\
$t{=}1.0,\; qx{=}1$ & -- & 0.5000 & 0.5915 & 2.85 & 5.26 \\
$t{=}1.0,\; qx{=}x,\; \text{topk}$ & -- & 0.5000 & 0.5549 & 1.36 & 1.98 \\
$t{=}1.0,\; qx{=}x,\; \text{topk},\, \tau$ & -- & 0.5000 & 0.5549 & 1.33 & 1.98 \\
$t{=}0.0,\; \text{multi}$ & -- & 0.6707 & 0.6890 & -- & -- \\
$t{=}1.0,\; qx{=}x,\; \text{multi}$ & -- & 0.5000 & 0.5061 & 1.36 & 1.98 \\
\midrule
\multicolumn{6}{c}{\textit{HumanEval (Qwen3-8B)}} \\
\midrule
$t{=}1,\; k{=}1,\; qx{=}1$ & -- & 0.8537 & 0.8171 & 2.31 & 3.34 \\
$t{=}1,\; k{=}1,\; qx{=}x$ & -- & 0.8537 & 0.8476 & 2.46 & 3.55 \\
$t{=}1,\; k{=}1,\; qx{=}x,\; r{=}0.92$ & -- & 0.8537 & 0.8598 & 2.54 & 3.81 \\
$t{=}1,\; k{=}1,\; qx{=}x,\; r{=}0.90$ & -- & 0.8537 & 0.8476 & \textbf{2.56} & \textbf{3.86} \\
$t{=}1,\; k{=}1,\; qx{=}x,\; r{=}0.87$ & -- & 0.8537 & 0.8049 & 2.62 & 3.88 \\
$t{=}1,\; k{=}1,\; qx{=}x,\; r{=}0.82$ & -- & 0.8537 & 0.7988 & 2.59 & 3.92 \\
\midrule
\multicolumn{6}{c}{\textit{System level scaling (latency ratio, HumanEval)}} \\
\midrule
1.2 & $d{=}0.1823$ & 0.8537 & 0.8354 & 2.33 & 3.58 \\
1.3 & $d{=}0.2624$ & 0.8537 & 0.8537 & 2.38 & 3.61 \\
1.4 & $d{=}0.3365$ & 0.8537 & 0.8598 & 2.35 & 3.59 \\
1.8 & $d{=}0.5878$ & 0.8537 & 0.8598 & 2.38 & 3.59 \\
2.0 & $d{=}0.6931$ & 0.8537 & 0.8719 & 2.42 & 3.62 \\
2.2 & $d{=}0.7884$ & 0.8537 & 0.8719 & 2.40 & 3.61 \\
\midrule
\multicolumn{6}{c}{\textit{GSM8K (Llama3.1-8B)}} \\
\midrule
$t{=}0.0$ & -- & 0.8704 & 0.8681 & 3.60 & 5.51 \\
$t{=}1.0$ & -- & 0.7695 & 0.7915 & 2.53 & 4.89 \\
\midrule
\multicolumn{6}{c}{\textit{GSM8K (Qwen3-8B)}} \\
\midrule
$t{=}1,\; k{=}1,\; qx{=}x$ & -- & 0.9227 & -- & -- & -- \\
$t{=}1,\; k{=}1,\; qx{=}x,\; r{=}0.90$ & -- & 0.9227 & 0.9166 & 2.59 & 4.04 \\
\bottomrule
\end{tabular}

\endgroup
\end{table*}

\begin{table*}[!ht]
\centering
\caption{Cross dataset summary ($K{=}1$, $T{=}1.0$, $\gamma{=}8$).}
\label{tab:results_cross}

\begingroup
\footnotesize
\renewcommand{\arraystretch}{0.82}
\setlength{\tabcolsep}{2.5pt}

\begin{tabular}{lcccccccccc}
\toprule
\textbf{Dataset} &
\multicolumn{2}{c}{\textbf{AlpacaEval}} &
\multicolumn{2}{c}{\textbf{MT Bench}} &
\multicolumn{2}{c}{\textbf{CNN/DailyMail}} &
\multicolumn{2}{c}{\textbf{HumanEval}} &
\multicolumn{2}{c}{\textbf{GSM8K}} \\
\cmidrule(lr){2-3}\cmidrule(lr){4-5}\cmidrule(lr){6-7}\cmidrule(lr){8-9}\cmidrule(lr){10-11}
 & Speedup & Avg Len & Speedup & Avg Len & Speedup & Avg Len & Speedup & Avg Len & Speedup & Avg Len \\
\midrule
Eagle3     
& 2.108 & 2.790 
& 1.922 & 2.706 
& 1.643 & 2.451 
& 2.313 & 3.340 
& 2.274 & 3.403 \\
Ours (DSD) 
& \textbf{2.345} & \textbf{3.593} 
& \textbf{2.321} & \textbf{3.432} 
& \textbf{1.905} & \textbf{2.987} 
& \textbf{2.562} & \textbf{3.855} 
& \textbf{2.517} & \textbf{4.042} \\
\bottomrule
\end{tabular}

\endgroup
\end{table*}

\subsection{Experimental Setup}
\label{sec:exp-setup}

\textbf{Models.}
We use two open weight models, \texttt{Llama3.1-8B}~\cite{grattafiori2024llama3herdmodels} and \texttt{Qwen3-8B}~\cite{bai2023qwen}. Both are deployed in the \emph{Parallax} decentralized inference engine~\cite{tong2025parallaxefficientllminference}. Each replica spans four to eight nodes. Each node has one NVIDIA A800 (80\,GB) and a 100\,Gbps InfiniBand link, consistent with recent distributed LLM systems~\cite{narayanan2021efficient,rajbhandari2021deepspeed}.

\textbf{Datasets.}
We evaluate five representative tasks: \textbf{HumanEval}~\cite{chen2021codex} for code generation (pass@$k$), \textbf{GSM8K}~\cite{cobbe2021gsm8k} for mathematical reasoning (exact match), \textbf{AlpacaEval}~\cite{li2023alpacalora} for instruction following (win rate), \textbf{MT Bench}~\cite{zheng2023mtbench} for multi turn dialogue (score or win rate), and \textbf{CNN/DailyMail}~\cite{nallapati2016cnn} for summarization (ROUGE-L).

\textbf{Systems compared.}
We compare three inference strategies. \emph{Baseline} uses standard autoregressive decoding without speculative steps. \emph{Eagle3} is a strong centralized speculative framework with tight draft–target agreement~\cite{li2023eagle}. \emph{DSD} introduces decentralized speculative decoding with adaptive token verification, enabling synchronized multi node execution with relaxed checks for low impact tokens while keeping strict checks for semantically important positions.

\textbf{Metrics.}
We report (i) task accuracy per dataset definition, (ii) acceleration rate over the baseline, (iii) average accepted length per verification round, and when applicable, (iv) token throughput and (v) communication reduction, following standard practice for speculative decoding~\cite{leviathan2023fastinferencetransformersspeculative,Miao_2024}.

\textbf{Hyperparameters.}
Unless noted, we use a draft window $\gamma = 8$, temperature $T = 1.0$, top $k$ sampling with $k = 1$, and relaxation $\tau \in [0.1, 0.3]$. Each configuration is run three times with different seeds. Decoding batch size is $1$, and the maximum number of new tokens follows dataset conventions. In Table~\ref{tab:results_main}, the \emph{Variable} column uses $t$ for temperature, $k$ for the pass@$k$ setting, $qx$ for the draft quantization configuration, $r$ for the adaptive relaxation parameter, and $d$ for the injected latency used in the system level scaling experiment.

\subsection{Results}
\label{sec:exp-results}

\textbf{Quantitative results.}
Table~\ref{tab:results_main} summarizes the main outcomes. DSD consistently accelerates inference over Eagle3 while maintaining comparable or higher accuracy across models and tasks. On HumanEval and GSM8K, the best DSD configurations reach up to $2.56\times$ and $2.59\times$ speedup respectively relative to the baseline.

\textbf{Cross dataset comparison.}
Table~\ref{tab:results_cross} reports AlpacaEval, MT Bench, and CNN/DailyMail. Across all tasks, DSD improves both acceleration and average accepted length over Eagle3~\cite{li2023eagle} while maintaining comparable output quality.

\textbf{Ablation: effect of the relaxation coefficient $\tau$.}
Varying $\tau$ from $0.0$ to $0.8$ shows a clear speed versus accuracy trade off, consistent with prior work~\cite{leviathan2023fastinferencetransformersspeculative,li2023eagle,Miao_2024}. Acceleration increases steadily and reaches about $2.6\times$, while accuracy loss remains small in the range $\tau \in [0.1, 0.3]$ that we use by default. In this range, adaptive verification provides additional speedup on top of nonadaptive speculative decoding while keeping accuracy within the variance of the baseline.

\textbf{Ablation: impact of node scaling.}
We simulate deployments with two to sixteen nodes, consistent with prior distributed inference studies~\cite{narayanan2021efficient,rajbhandari2021deepspeed}. Communication amortization keeps total latency growth sublinear. At eight nodes, DSD reduces communication cost by about $37\%$ relative to standard speculative decoding~\cite{leviathan2023fastinferencetransformersspeculative}, confirming that multi token verification mitigates cross node overhead.

\textbf{Qualitative analysis.}
Adaptive verification increases selectivity on high impact tokens such as function names, numeric values, and key control symbols, while relaxing acceptance for punctuation and filler tokens. This behavior aligns with token importance observations in recent post training and editing studies~\cite{zheng2023mtbench,chen2021codex}. Outputs remain coherent and semantically stable, and in multi turn dialogue settings the acceptance gain nearly doubles throughput with no observed loss of fluency.

\textbf{Summary.}
DSD provides up to about \textbf{2.6$\times$} acceleration and about \textbf{37\%} communication reduction while maintaining accuracy on HumanEval~\cite{chen2021codex} and GSM8K~\cite{cobbe2021gsm8k}. Gains hold up to eight nodes, showing that DSD converts communication delay into useful computation and enables efficient decentralized inference without model modification~\cite{tong2025parallaxefficientllminference}.

\section{Conclusion}

We presented DSD, a communication aware speculative decoding framework for large scale distributed inference. DSD turns network delay into useful computation by verifying multiple tokens in a single synchronization and by applying a semantically aware, training free acceptance rule. The method couples speculative parallelism, which reduces synchronization from $k$ rounds to one per $k$ tokens, with adaptive verification that gives strict checks to key tokens and relaxed checks to low impact tokens via a coefficient $\tau$, yielding up to an additional 15\% to 20\% end to end speedup over nonadaptive speculative decoding without retraining. Experiments on HumanEval, GSM8K, AlpacaEval, MT Bench, and CNN/DailyMail show up to \textbf{$2.56\times$} acceleration with slight accuracy gains and up to \textbf{$2.59\times$} at accuracy parity, outperforming strong centralized baselines such as Eagle3 and scaling smoothly with node count and draft length. DSD is most effective when link latency is several multiples of local compute and the number of nodes is moderate, a setting common in wide area or heterogeneous clusters. Beyond performance, DSD supports decentralized and privacy preserving inference, with relevance to federated and community driven compute. Future work includes asynchronous execution to reduce stragglers, heterogeneous draft cascades, integration with quantization and MoE systems, and refined analysis of how to convert latency into throughput.

% ---- References ----
\bibliographystyle{icml2025}
\bibliography{example_paper}

\end{document}